\documentclass{aa}

\usepackage{graphicx,amssymb,natbib}
\usepackage{txfonts}
\usepackage{mathrsfs}

\begin{document}

\def\A{\,\AA\ }
\def\AF{\,\AA}
\def\arcsec{\hbox{$^{\prime\prime}$}}
\newcommand{\etal}{et~al. }
\newcommand{\ca}{\ion{Ca}{ii} 8542~\A }
\newcommand{\fca}{\ion{Ca}{ii} 8542~\AA}
\newcommand{\cmv}{\mbox{{\rm\thinspace cm$^{-3}$}}}
\newcommand{\cmmf}{\mbox{{\rm\thinspace cm$^{-5}$}} }
\newcommand{\D}{\displaystyle}
\newcommand{\eq}[1]{Eq.\,(\ref{#1})}
\newcommand{\fig}[1]{Fig.~\ref{#1}}
\newcommand{\km}{\mbox{{\rm\thinspace km}}}
\newcommand{\kms}{\mbox{{\rm\thinspace km\thinspace s$^{-1}$}}}
\newcommand{\ms}{\mbox{{\rm\thinspace m\thinspace s$^{-1}$}}}
\newcommand{\K}{\mbox{{\thinspace\rm K}}}
\newcommand{\tab}[1]{Table~\ref{#1}}

\newcommand{\ben}{\begin{enumerate}}
\newcommand{\een}{\end{enumerate}}
\newcommand{\bfig}{\begin{figure}}
\newcommand{\efig}{\end{figure}}
\newcommand{\beq}{\begin{equation}}
\newcommand{\eeq}{\end{equation}}
\newcommand{\mbf}{\mathbf}
\newcommand{\vare}{\varepsilon}
\newcommand{\mcal}{\mathcal}
\newcommand{\ep}{\epsilon}
\newcommand{\cs}{\mathcal{S}}
\newcommand{\csv}{\mathcal{S}_{\mathcal{V}}}
\newcommand{\cv}{\mathcal{V}}
\renewcommand{\thefootnote}{\dag}
\def\degr{\hbox{$^{\circ}$}}

\newcommand{\den}{$N_{\rm e}$ }
\newcommand{\fden}{$N_{\rm e}$}
\newcommand{\mic}{$\xi_{\rm t}$ }
\newcommand{\fmic}{$\xi_{\rm t}$}
\newcommand{\Ha}{${\rm H\alpha}$ }
\newcommand{\ha}{${\rm H\alpha}$ }
\newcommand{\fha}{${\rm H\alpha}$}
\newcommand{\haw}{$\pm$0.29\A}
\newcommand{\caw}{$\pm$0.09\A}
\newcommand{\hawt}{$\pm$0.58\A}
\newcommand{\cawt}{$\pm$0.135\A}
\def\degr{\hbox{$^{\circ}$}}

\title{Energy and helicity budgets of solar quiet regions}

\author{K.\,Tziotziou\inst{1,2}
        \and G.\,Tsiropoula\inst{2} \and M.K.\, Georgoulis\inst{1}
        \and I.\,Kontogiannis\inst{2}}

\offprints{K.\,Tziotziou,\\
\email{kostas@noa.gr}}

\institute{Research Center for Astronomy and Applied Mathematics, Academy of
Athens, 4 Soranou Efesiou Street, Athens GR-11527, Greece \and
Institute for Astronomy, Astrophysics, Space Applications and Remote Sensing,
National Observatory of Athens,
GR-15236 Penteli, Greece}

\date{Received  / Accepted }

\titlerunning{Energy and helicity budgets in quiet Sun}

\authorrunning{Tziotziou \etal}

\abstract{}{We investigate the free magnetic energy and relative magnetic helicity budgets of solar quiet regions.}
{Using a novel non-linear force-free method requiring single solar vector magnetograms we calculate the instantaneous free
magnetic energy and relative magnetic helicity budgets in 55 quiet-Sun vector magnetograms.}
{As in a previous work on active regions, we construct here for the first time the (free) energy-(relative) helicity diagram of quiet-Sun regions. We find that quiet-Sun regions have no dominant sense of helicity and show monotonic correlations a) between free magnetic energy/relative helicity and magnetic network area and, consequently, b) between free magnetic energy and helicity. Free magnetic energy budgets of quiet-Sun regions represent a rather continuous extension of respective active-region budgets towards lower values, but the corresponding helicity transition is discontinuous due to the incoherence of the helicity sense contrary to active regions. We further estimate the instantaneous free magnetic-energy and relative magnetic-helicity budgets of the entire quiet Sun, as well as the respective budgets over an entire solar cycle.}
{Derived instantaneous free magnetic energy budgets and, to a lesser extent, relative magnetic helicity budgets over the entire quiet Sun are comparable to the respective budgets of a sizeable active region, while total budgets within a solar cycle are found higher than previously reported. Free-energy budgets are comparable to the energy needed to power fine-scale structures residing at the network, such as mottles and spicules.}


\maketitle

\section{Introduction}

Free magnetic energy corresponds to the excess energy of any magnetic region from its ``ground'',
current-free (potential) energy state, while magnetic helicity
quantifies the stress and distortion of the magnetic field lines compared
to their potential-energy state. Free magnetic energy builds up mainly through continuous flux emergence on the solar surface and other
processes such as coronal interactions \citep[e.g. ``fly-bys'',][]{gals00} or photospheric twisting \citep[e.g.][]{pari09}.
Magnetic helicity either emerges from the solar interior via helical magnetic flux tubes or is being generated
by solar differential rotation and peculiar photospheric motions.

In solar active regions (ARs) considerable, localized magnetic flux emergence of
the order of 10$^{22}$ Mx \citep{schr:harv} builds up strong opposite-polarity regions that are sometimes separated by
intense, highly sheared polarity inversion lines (PILs), hence deviating substantially from a potential-field configuration.
ARs tend to store large amounts of both free magnetic energy and magnetic helicity.
Free magnetic energy is released, via magnetic reconnection events, in solar flares and/or coronal mass ejections (CMEs). Helicity, however,
cannot be efficiently removed by magnetic reconnection \citep{berg84}, and if not transferred to larger scales via
existing magnetic connections, it can only be expelled in the form of CMEs \citep{low94, devo:00}.
The role of both free magnetic energy and helicity in ARs has been recently investigated by \citet{geo12a} and \citet{tzio12,tzio13}.

On the other hand, quiet-Sun regions are dominated by the flow pattern of large convective cells called supergranules that range in diameter
from 10\,000 km to 50\,000 km, with an average diameter between 13\,000 and 18\,000 km \citep{hage:97}.
High-resolution magnetograms show continuous emergence of new bipolar elements inside the cell interiors, called the internetwork (IN),
that are swept by the supergranular flow towards the boundaries of supergranular cells where opposite polarity fluxes cancel,
whereas like-polarity fluxes merge \citep{wang96,schr97}. By this process hierarchic flux concentrations are formed at the intersection of supergranular cells. These magnetic flux concentrations, which are characterized by magnetic fluxes of the order of 10$^{18}$-10$^{19}$ Mx and typical diameters of 1\,000-10\,000 km \citep{parn:01}, constitute the so called {\em magnetic network}. Free magnetic energy, released in the network mainly by reconnection can fuel the dynamics of several small-scale structures, such as mottles/spicules, residing there and governed by the dynamics and physics of the network magnetic field \citep[see the review by][for further details]{tsir:12}. Free magnetic energy release from non-potential magnetic configurations in the quiet Sun has also been reported to result in small-scale structures, such as bright points \citep{zhao09}, blinkers \citep{wood99} and quiet-Sun corona nanoflares \citep{meyer13}. There are no reports in the literature concerning the accumulation and expulsion of relative magnetic helicity in the magnetic network and quiet-Sun regions in general and this mechanism's role in quiet-Sun dynamics; only \citet{zhao09} have investigated current helicity budgets in network bright points. However, there exist reports \citep{jess09,curd11,depo12} arguing for torsional oscillations in fine structures,  such as explosive events and spicules, thus suggesting the existence of twisting motions that could lead to expulsion or transfer of helicity to larger scales in the solar atmosphere.

Until recently, no robust method existed to calculate the instantaneous free magnetic
energy and relative magnetic helicity budgets of a solar region. Existing methods were based
either on integration in time of an energy/helicity injection rate \citep{berfie84} or
evaluation of classical formulas \citep[hereafter {\em volume calculations},][]{finn85, berg99} using a
three-dimensional magnetic field derived from extrapolations. However, energy/helicity
injection rates depend on the determination of the photospheric
velocity field, which involves significant uncertainties \citep[e.g.,][]{wels07}. On the other hand,
volume calculations depend on model-dependent nonlinear force-free
(NLFF) field extrapolations that also carry several uncertainties and ambiguities
\citep[e.g.,][ and references therein]{schr06,metc08}. Recently, \citet{geo12a} proposed
a novel NLFF method to calculate the instantaneous magnetic free energy and relative helicity budgets
from a single (photospheric or chromospheric) vector magnetogram. The method was used for calculating
the free magnetic energy and relative helicity budgets in solar ARs \citep{geo12a,tzio13} and for
deriving the energy-helicity diagram of solar ARs \citep{tzio12}. The latter shows a nearly monotonic dependence
between the two quantities.

The aim of this paper, is to a) derive the instantaneous budgets of free magnetic energy and relative magnetic helicity
in quiet-Sun regions using the aforementioned NLFF method, b) construct the corresponding energy-helicity diagram
and compare it with the respective diagram for ARs, and c) calculate available budgets of free energy and
helicity over an entire solar cycle and associate them with the energetics and dynamics of fine-scale quiet-Sun structures.
Section~\ref{obsmeth} briefly describes and discusses the observations and the methodology, Sect.~\ref{res}
presents the results, while Sect.~\ref{conc} discusses the results and summarizes our findings.

\section{Observations and Methodology}
\label{obsmeth}

\subsection{Observations}
\label{obs}

For our analysis we use a sizable sample of photospheric
vector magnetograms of quiet-Sun regions obtained by the space-borne
Spectropolarimeter \citep[SP; see description in][]{lite08}
of the Solar Optical Telescope (SOT) onboard the {\em Hinode} satellite. SP provides
full Stokes profiles of the Fe I 630.25/630.15 nm lines with a maximum spatial sampling of
0.16\arcsec\ per pixel. All vector magnetograms were treated for the azimuthal 180\degr\ ambiguity, by means of
the non-potential field calculation (NPFC) of \citet{geo05}, as revised in \citet{metc06}. For our analysis
we use the heliographic components of the magnetic field vector on the heliographic plane,
derived with the de-projection equations of \citet{gary90}. As typical uncertainties for the line-of-sight and
transverse field components of SOT/SP data we use ($\delta$B$_l$, $\delta$B$_{tr}$)=(5, 50) G.

\begin{figure}
  \includegraphics[width=\linewidth, bb = 112 95 517 471]{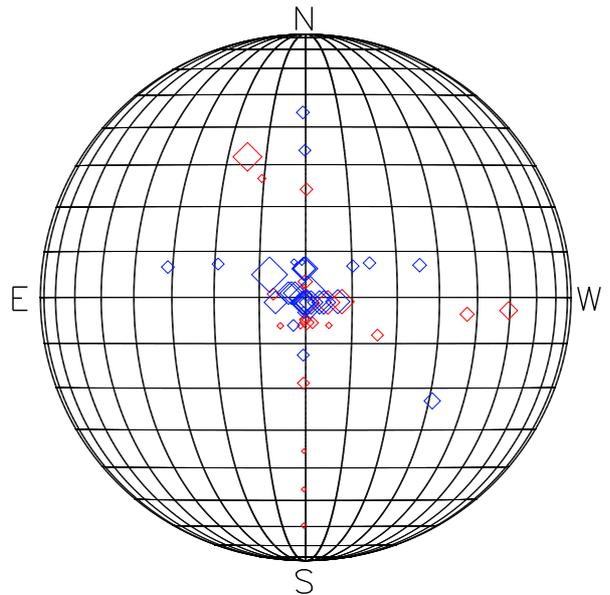}
     \caption{Central heliographic positions for the 55 vector quiet Sun magnetograms included in our analysis.
     The size of the diamonds implies normalized (to the maximum value) area size. Red/blue diamonds
     correspond to negative/positive total magnetic helicity budgets (see Section~\ref{enhelbudgets}).}
     \label{noaapos}
\end{figure}

A total of 132 quiet Sun regions, observed between 2006 and 2011 when the
Sun was mostly quiet, were initially examined. Of them, we selected 55 magnetograms that exhibited an overall flux imbalance not
exceeding 15\%, with an average imbalance of the order of 5\%. This is because our NLFF calculation gives more reliable results in
flux-balanced environments. The spatial resolution of the analyzed magnetograms is 0.16\arcsec\ or 0.32\arcsec\ per pixel and
Fig.~\ref{noaapos} shows the spatial distribution of these regions on the solar disc. The majority of them is located close to the solar disc center and cover a sizeable area of the solar surface; areas on the image plane range between 1\,426 and 86\,088 arcsec$^2$, with a mean area size of 32\,078 arcsec$^2$.

\begin{figure}
  \includegraphics[width=\linewidth]{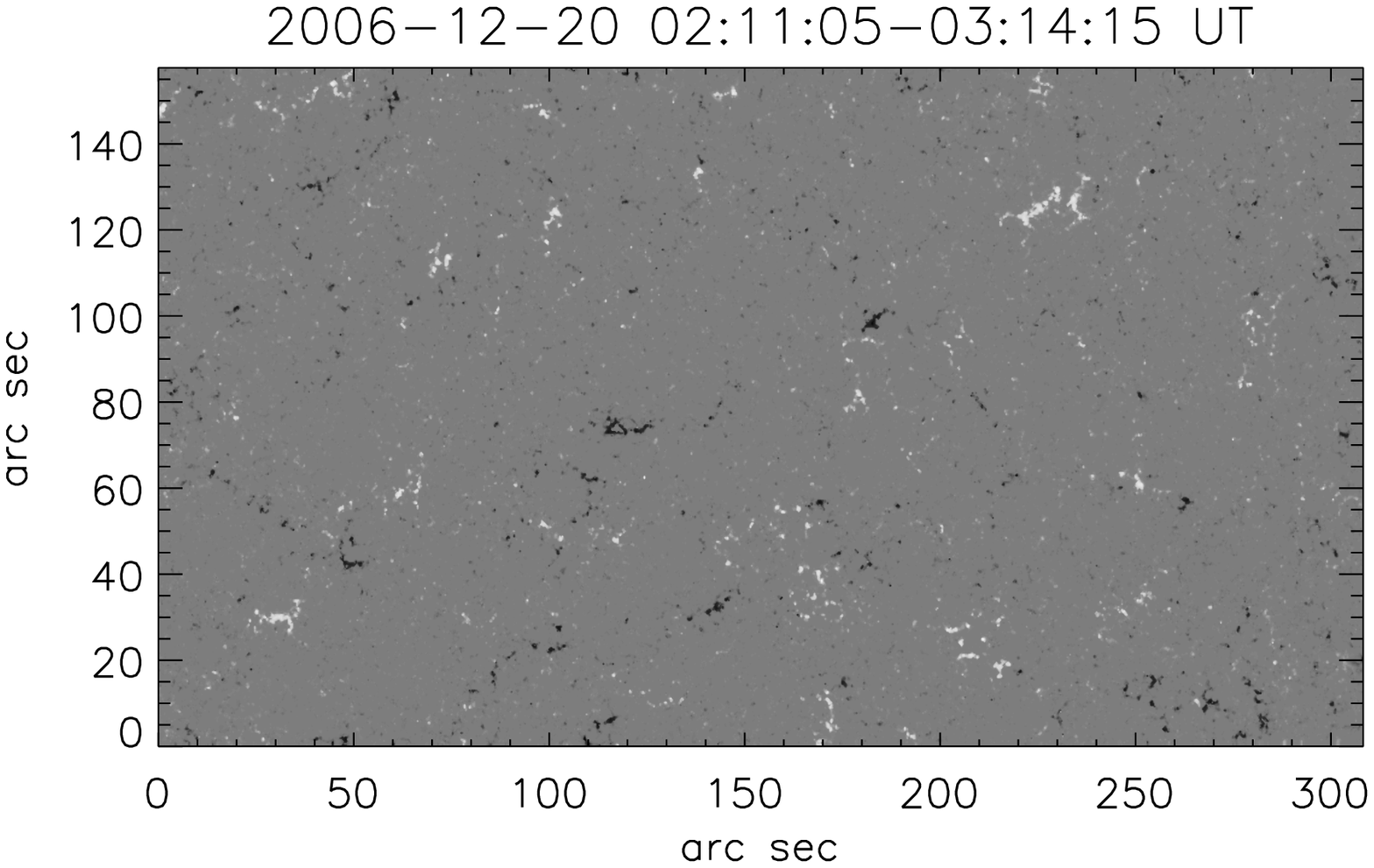}\\
  \includegraphics[width=\linewidth]{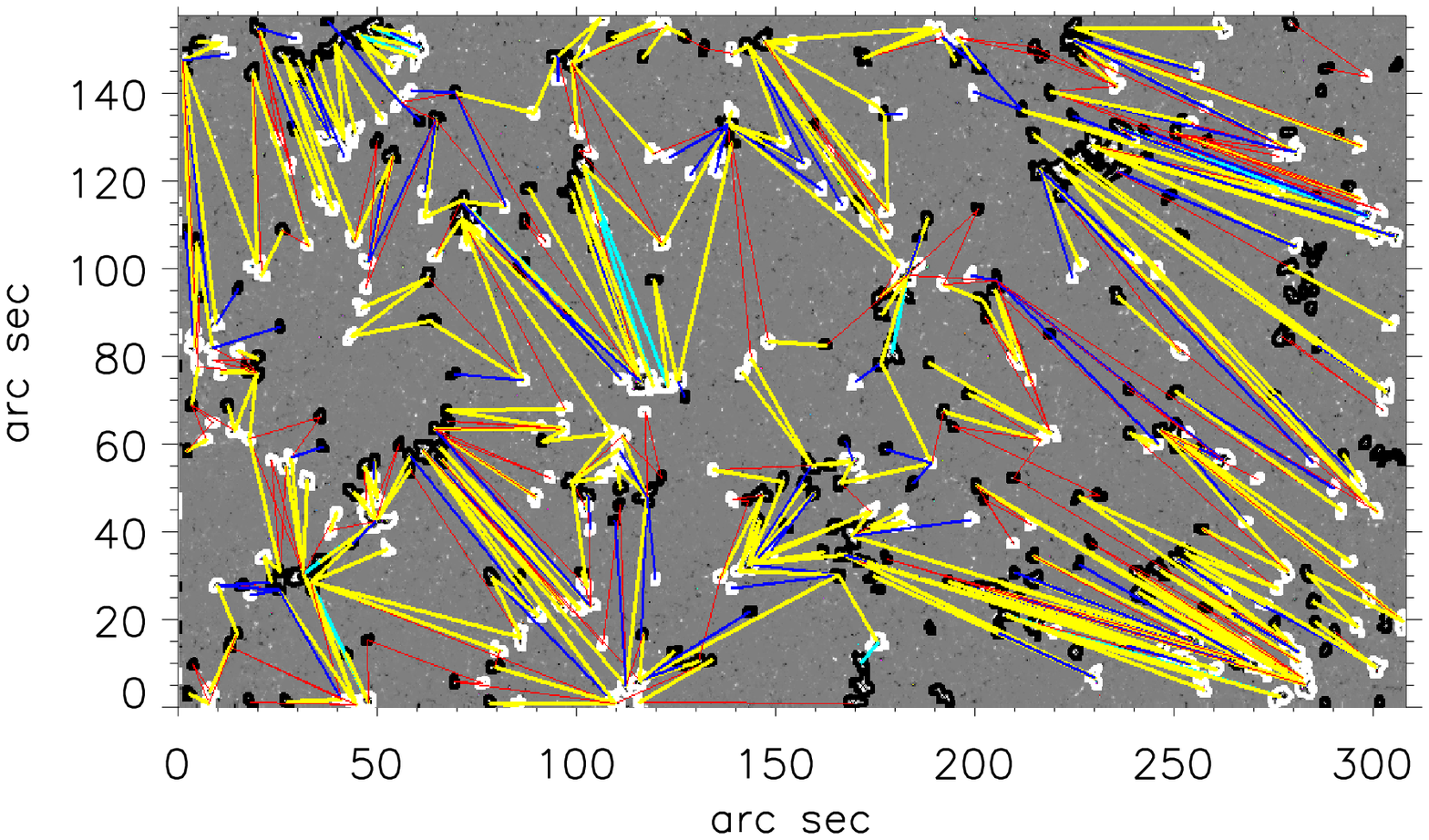}
     \caption{Top panel: An observed SOT/SP magnetogram (vertical field component B$_{\rm z}$) of a quiet-Sun region studied here. White/black denotes
     positive/negative magnetic field concentrations. Bottom panel: Inferred magnetic connectivity of the region, showing
      the vertical magnetic field component in grayscale with the contours bounding the identified magnetic partitions.
      The flux-tube connections identified by the magnetic connectivity matrix are represented by line segments connecting the flux-weighted centroids
      of the respective pair of partitions. Red, blue, yellow, and cyan segments denote magnetic flux
      contents within the ranges [10$^{16}$, 10$^{17}$] Mx, [10$^{17}$,10$^{18}$] Mx, [10$^{18}$, 10$^{19}$] Mx, and >10$^{19}$ Mx,
      respectively. Only connections closing within the field-of-view are shown. }
     \label{quietreg}
\end{figure}

Figure~\ref{quietreg} (top panel) shows an example of a SOT/SP magnetogram used in this analysis. As the figure
shows, the magnetic field is mostly concentrated at the boundaries of the supergranular cells that constitute the magnetic network.

\subsection{Methodology for calculating magnetic energy and helicity budgets}
\label{newmethod}

The method for deriving the instantaneous free magnetic-energy and relative magnetic-helicity budgets is
a recently proposed NLFF method  by \citet{geo12a} that uses a single photospheric or chromospheric
vector magnetogram. Contrary to model-dependent NLFF field extrapolation methods, it provides unique results
relying on a unique magnetic-connectivity matrix that contains the flux committed
to connections between positive- and negative-polarity flux partitions. Since the three-dimensional
magnetic configuration is unknown, the aforementioned connectivity matrix is derived by means of a
simulated annealing method \citep{geo:rus,geo12a}. This method guarantees connections between opposite-polarity flux
partitions while globally (within the field-of-view) minimizing the corresponding connection lengths.
The non-zero flux elements of the derived connectivity matrix define a collection of $N$ magnetic
connections which are treated as slender force-free flux tubes with known footpoints,
flux contents, and variable force-free parameters.

For flux tubes that do not wind around each other's axes, as \citet{geo12a} have shown, the lower-limit free magnetic
energy $E_c$ can be expressed as the sum of two terms: a self term $E_{c_{\rm self}}$, describing the internal twist
and writhe of each flux tube, and a mutual term $E_{c_{\rm mut}}$, describing interactions
between different flux tubes. $E_c$ is defined by:
\begin{eqnarray}
E_c  & = & E_{c_{\rm self}} + E_{c_{\rm mut}} \nonumber \\
 & = & A d^2 \sum _{l=1}^N \alpha _l^2
\Phi_l^{2 \delta} +
      \frac{1}{8 \pi} \sum _{l=1}^N \sum _{m=1, l \ne m}^N
           \alpha _l \mcal{L}_{lm}^{\rm arch} \Phi_l \Phi_m\;\;,
\label{Ec_fin}
\end{eqnarray}
where $d$ is the pixel size of the magnetogram, $A$ and $\delta$ are known fitting
constants,  and $\Phi_l$ and $\alpha_l$ are the respective unsigned flux and
force-free parameters of flux tube $l$. $\mcal{L}_{lm}^{\rm arch}$ is the
mutual-helicity factor of two arch-like flux tubes which was first introduced by
\citet{dem06} and was further analyzed by \cite{geo12a} for all possible
cases. Since the winding factor (Gauss linking number) around flux tubes is unknown and hence set to zero,
the derived free energy always represents a lower, but realistic limit \citep{geo12a, tzio12}.

Likewise, the corresponding relative magnetic helicity $H_m$ is the sum of a
self $H_{m_{\rm self}}$ and a mutual $H_{m_{\rm mut}}$ term, namely
\begin{eqnarray}
H_m  & = & H_{m_{\rm self}} + H_{m_{\rm mut}} \nonumber \\
&  = & 8 \pi d^2 A
\sum _{l=1}^N \alpha _l \Phi_l ^{2 \delta} +
      \sum _{l=1}^N \sum _{m=1,l \ne m}^N \mcal{L}_{lm}^{\rm arch} \Phi_l
      \Phi_m\;\;.
\label{Hm_fin}
\end{eqnarray}
Derivation of uncertainties for all terms of Eqs. (\ref{Ec_fin}) and (\ref{Hm_fin}) is fully described in \citet{geo12a}.

Figure~\ref{quietreg} (bottom panel) shows the inferred connectivity matrix for the quiet-Sun region shown in the top panel.
Positive/negative partitions are, as expected, mostly concentrated in supergranular boundaries (magnetic network), where considerable
concentrations of magnetic flux are observed. Moreover, connections are mostly between nearby opposite-polarity partitions as
a result of the used simulated annealing method.

\subsubsection{Validity of the method}
\label{newmethodval}

\citet{geo12a} discussed in detail (see their Sect.~2.3) the validity of the aforementioned method for potential-field
configurations. The authors argued that non-potentiality of the studied region is a necessary assumption for the derivation
of the magnetic connectivity matrix with simulated annealing, since it strongly favors PILs. And if
non-potentiality characterizes solar active regions \citep[e.g.,][]{ziri:93,leka:96} this is not the case for quiet-Sun regions which are widely believed to be close to a potential-field configuration and have been mostly treated as such \citep[e.g.,][]{konto10,konto11,wieg13}. However, there exist observations and modeling suggesting the existence of non-potential fields within the quiet-Sun and its magnetic network \citep{wood99,zhao09,liu11,urit13,meyer13,ches13}.

Furthermore, there are a few elements characterizing the magnetic field configuration of the quiet Sun that would still allow use of simulated annealing. First, an observed quiet Sun magnetogram is usually not
flux-balanced. As a result, mutual-helicity terms could still sum up to give considerable helicity values when free
magnetic energy tends to zero due to the very small, close to zero, values of $\alpha _l$ in Eq.~(\ref{Ec_fin}), as
electric currents are almost absent. Indeed our results (see Figs.~\ref{hmecarea} and \ref{ehdiag} and relevant discussion) indicate that free magnetic energy acquires very low values compared to those for ARs.
Second, the quiet-Sun magnetic field concentrates on hierarchical structures (network). And even though there are no strong PILs present in
quiet Sun magnetograms, the supergranular cell configuration of the network favors an essential ingredient of the used simulated annealing, i.e.
simultaneous minimization of the flux imbalance and the separation length between chosen partitions, with the latter often being shorter than the supergranular cell radius (see Fig.~\ref{quietreg}).

\begin{figure*}
  \includegraphics[width=0.33\linewidth]{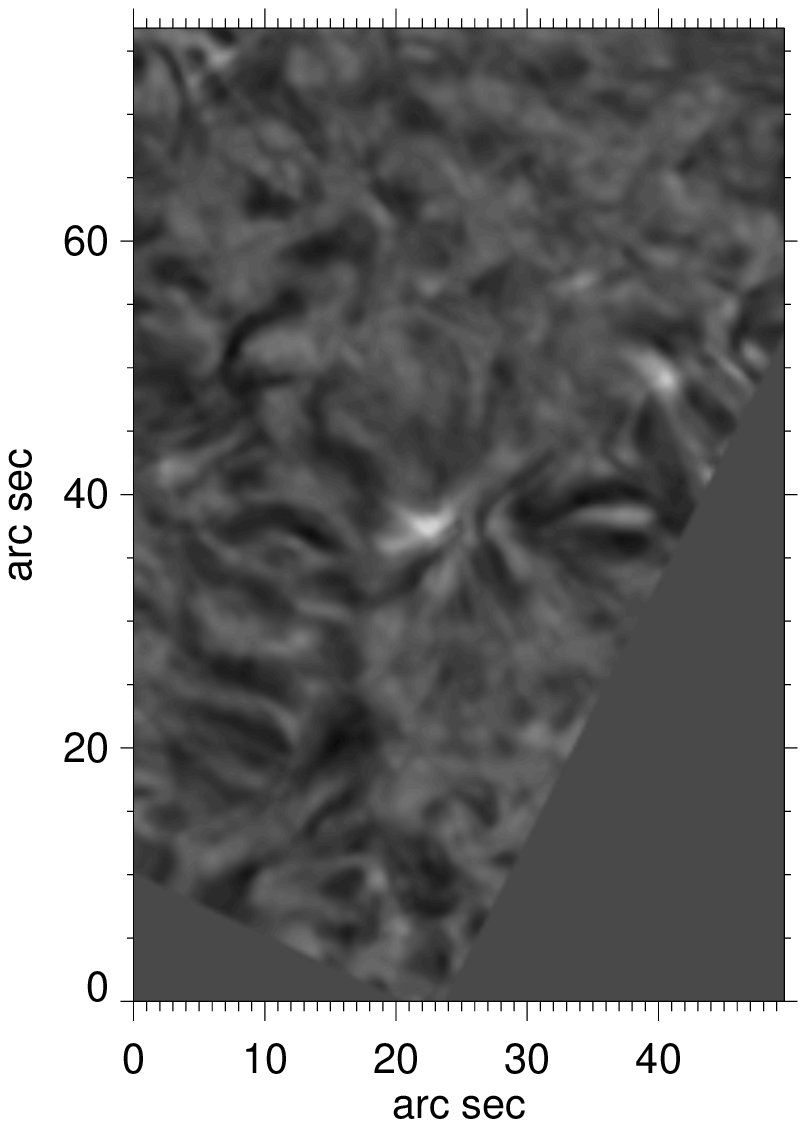}
  \includegraphics[width=0.33\linewidth]{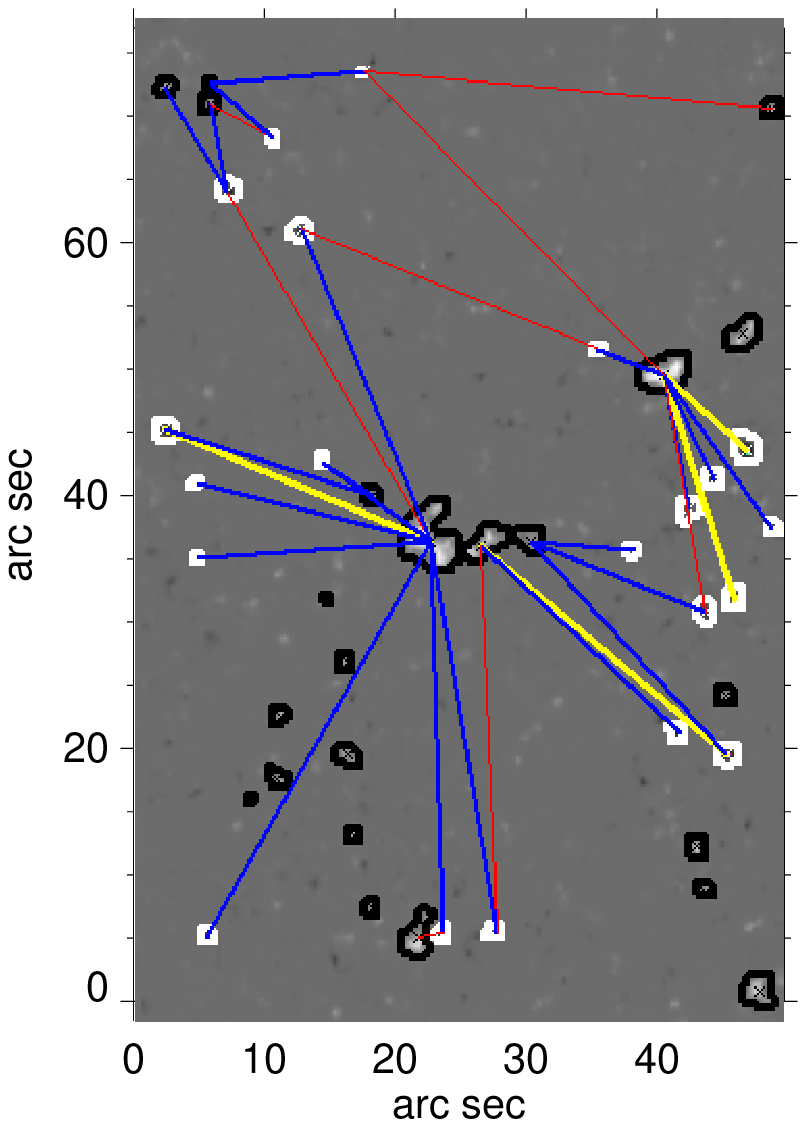}
  \includegraphics[width=0.33\linewidth]{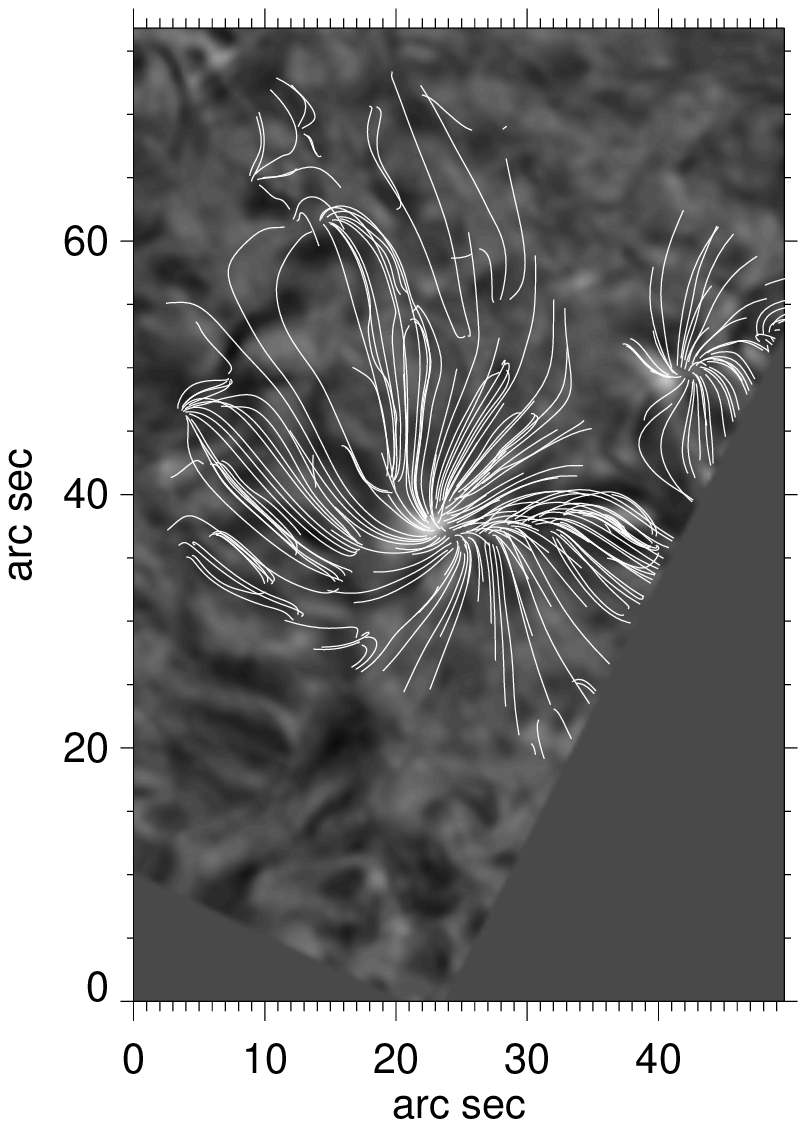}
     \caption{Left panel: An observed H$\alpha$ image observed with the Dutch Open Telescope \citep[see][for details]{konto10}. Middle panel: The corresponding SOT/SP magnetogram (vertical field component B$_{\rm z}$) of the quiet-Sun region with the inferred connectivity overlaid (see caption of Fig.~\ref{quietreg} for an explanation of the magnetic connectivity matrix). Right panel: The field lines of the potential field extrapolation of \citet{konto10} (their Fig.~2) overlayed on the H$\alpha$ image of the left panel.}
     \label{haimag}
\end{figure*}

Fig.~\ref{haimag} shows an H$\alpha$ quiet-Sun region (left panel) and its SOT/SP magnetogram (middle panel) that was studied by \citet{konto10}. They calculated the potential magnetic field of the chromosphere and in their Fig.~2 ({\em potential case} hereafter),
the magnetic field lines were plotted over an H$\alpha$ average image which is part of the one shown in Fig.~\ref{haimag}, rotated by 26\degr, and shown here for direct comparison in the right panel of Fig.~\ref{haimag}. This gives us the opportunity to compare between the magnetic connectivity as calculated for a potential magnetic field and by the annealing method used in this paper. This very quiet solar region is located almost at the disk center within 20\arcsec and 45\arcsec. Two major positive-polarity clusters stand out, that reach 1 kG, and some negative and positive ones, with lower strength (a few hundred G) while several mixed-polarity elements are scattered across the IN. We should however note that the signed magnetic flux is predominantly positive (45\% imbalance) and the annealing method takes into account only  partitions with fluxes above a certain threshold.

In the potential case, the two positive-polarity network clusters connect partly with the negative polarity elements at the middle-right of the field-of-view (FOV) and with several negative magnetic elements at the IN. The bundles of field lines coincide nicely with the absorption features in H$\alpha$. For the most part the same connectivity is inferred from our method (Fig.~\ref{haimag}, middle panel), but at a lower detail because the imposed threshold excludes very weak partitions. However, most of the connections (if not all) with the IN are lost. At the left of the cluster, parallel field lines connect weak opposite polarities and appear as a chain of mottles in H$\alpha$. This structure is not found in middle panel of  Fig.~\ref{haimag}: only the strongest negative partitions are detected and they are connected with the network cluster. Several small-scale flux tubes, formed at the IN, in the vicinity (or at places) of H$\alpha$ absorption, are also missed. Some very long potential magnetic field lines connect the network with remote magnetic elements; these connections are also found here and the inferred flux tubes match with the general orientation of absorption features, even though such long flux tubes are not usually observed in H$\alpha$.

As a concluding remark, our flux threshold, which is necessary for the calculation of the flux tubes that connect opposite-polarity elements, results in missing the finest magnetic fields of the internetwork. Moreover, several other possible connections are missing as a result of the non-flux balanced FOV. However, the method reproduces the general configuration inferred from H$\alpha$ observations.

\begin{figure*}
  \includegraphics[width=0.49\linewidth]{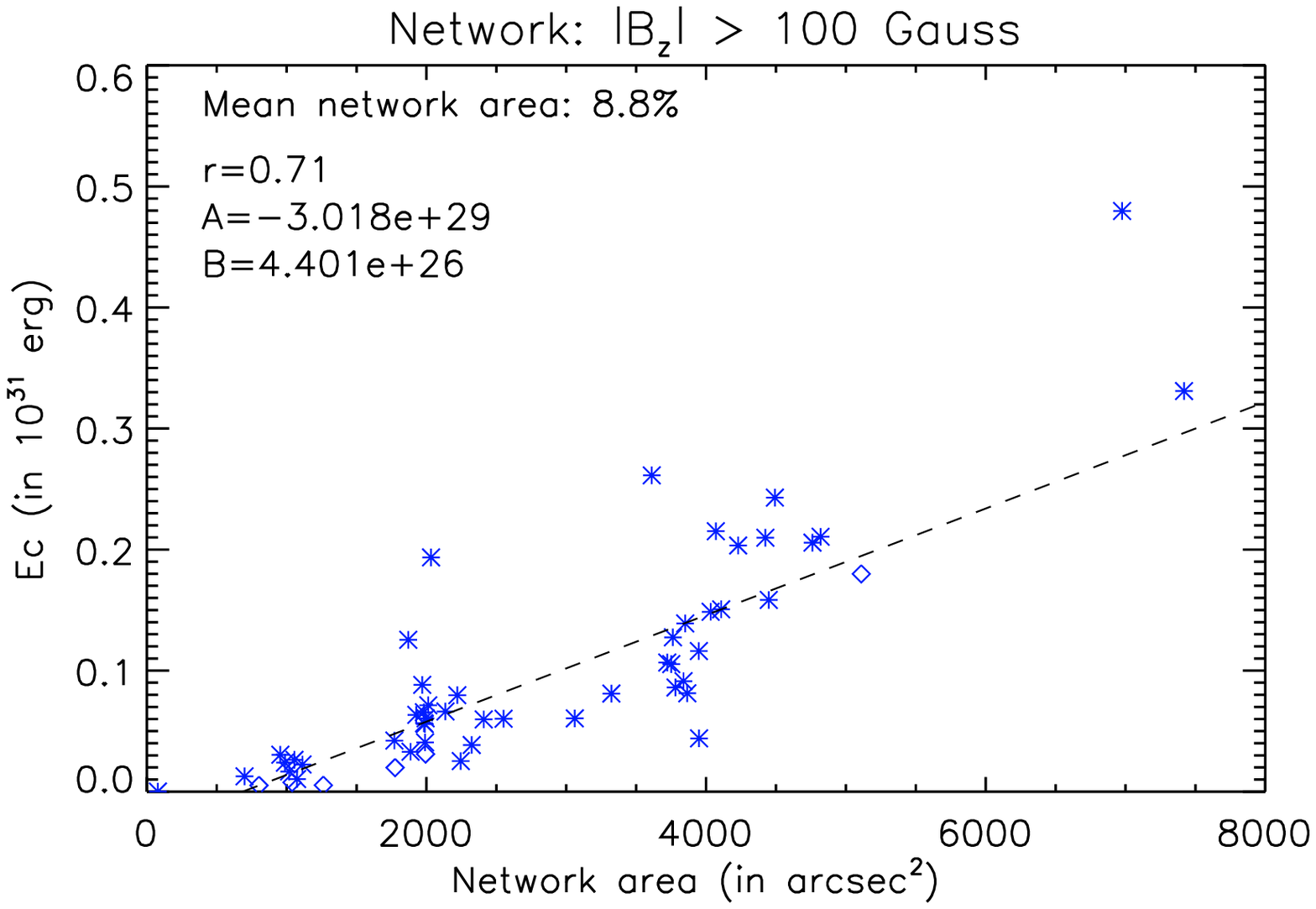}
  \includegraphics[width=0.49\linewidth]{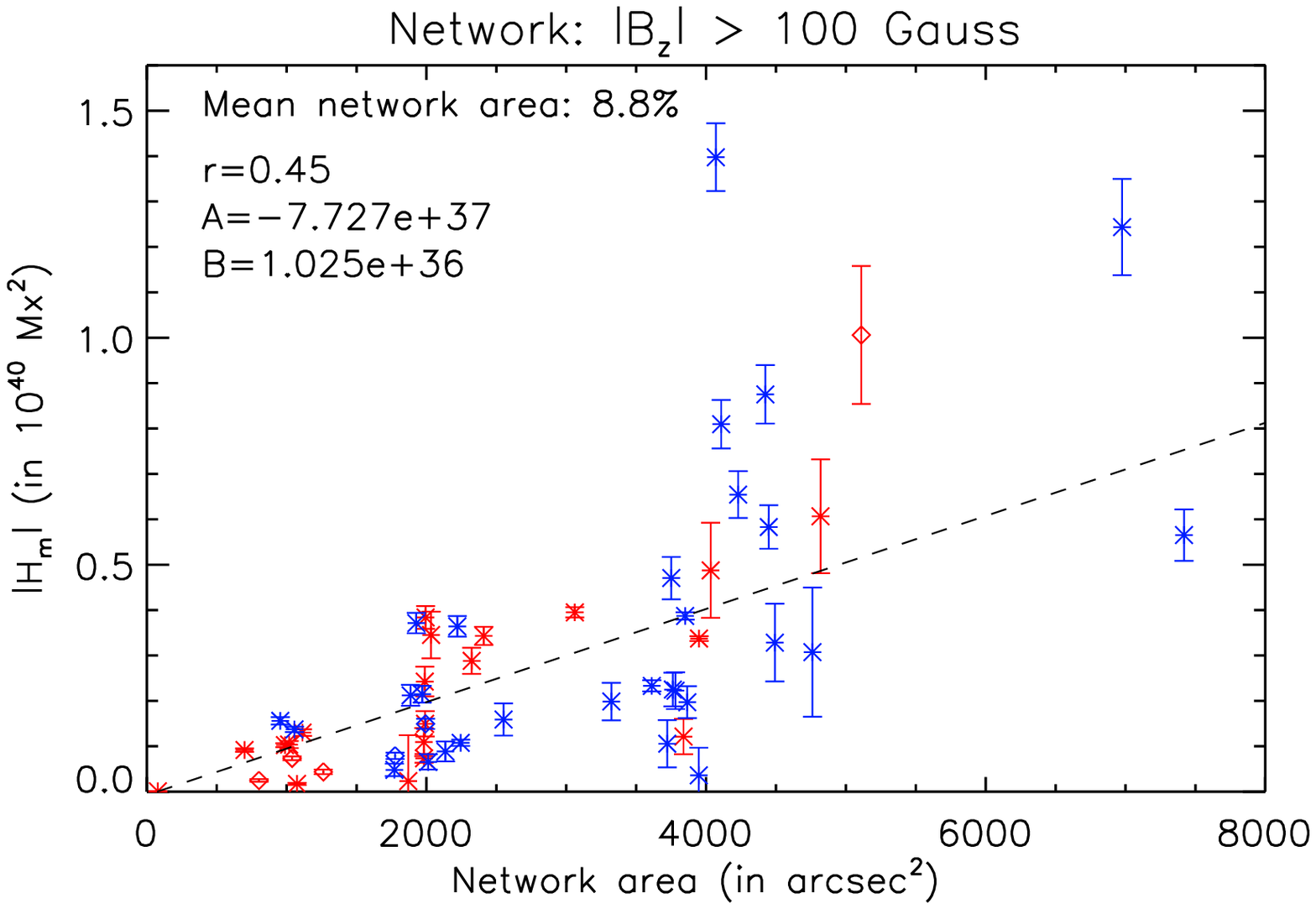}\\
  \includegraphics[width=0.49\linewidth]{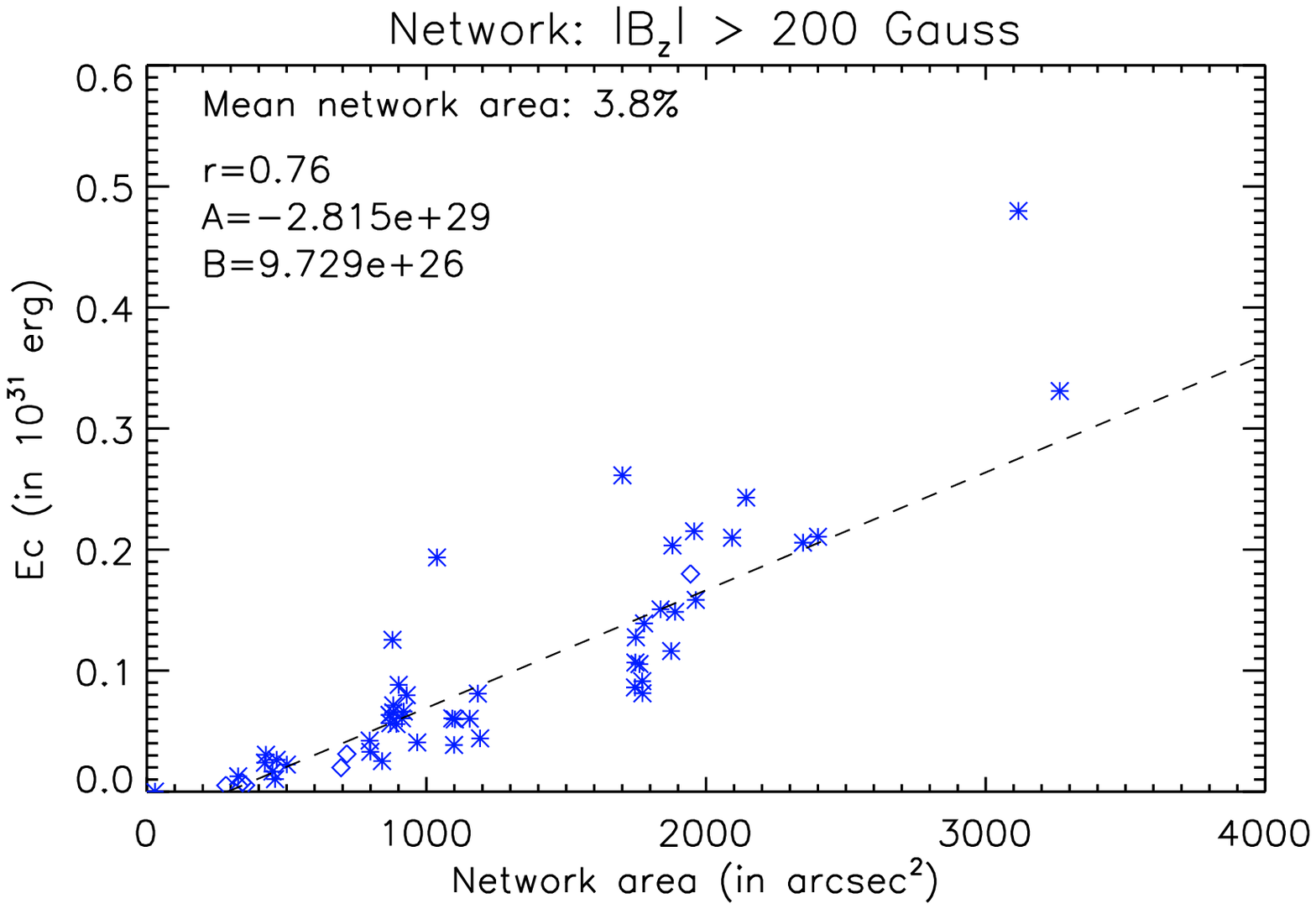}
  \includegraphics[width=0.49\linewidth]{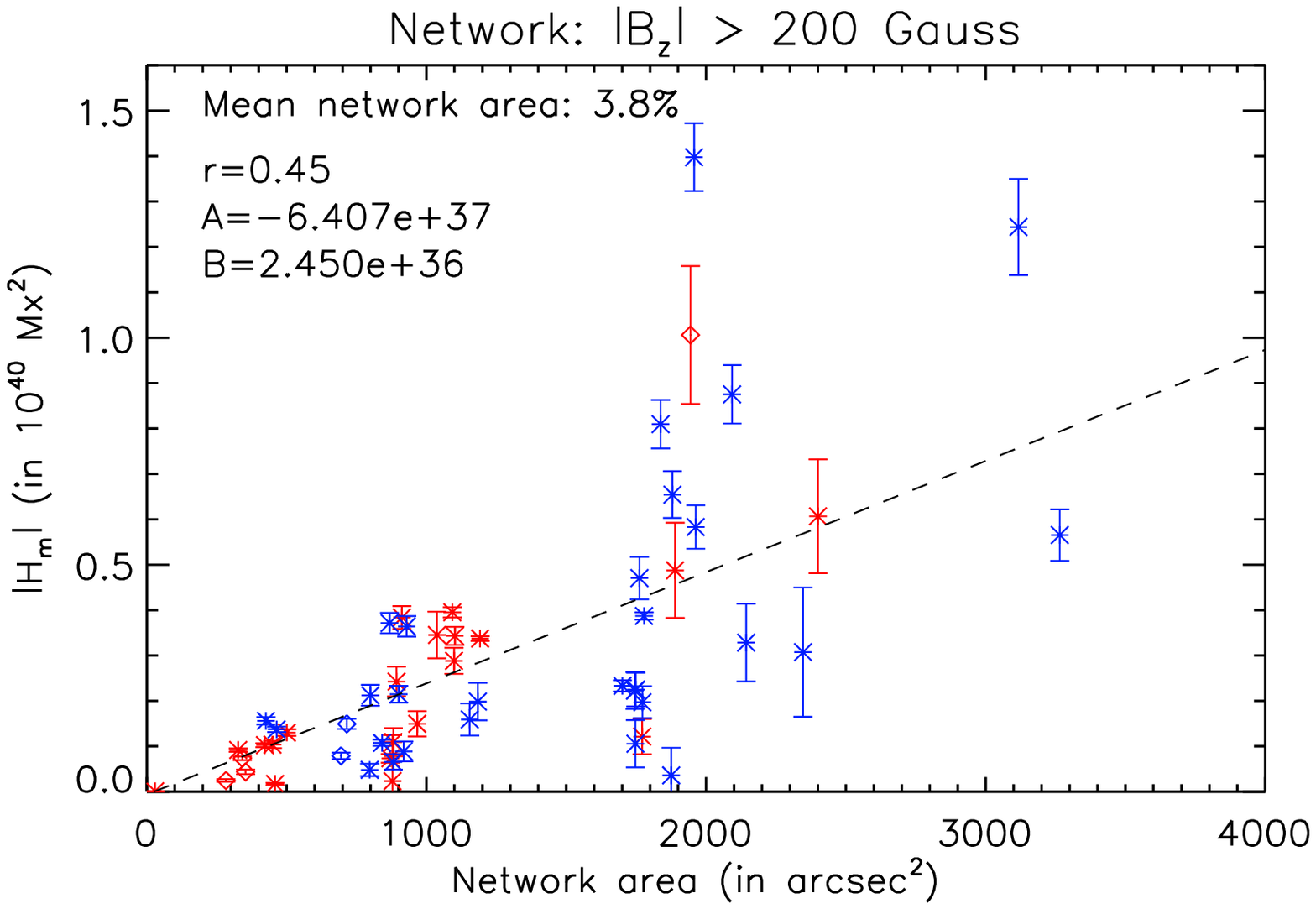}\\
  \includegraphics[width=0.49\linewidth]{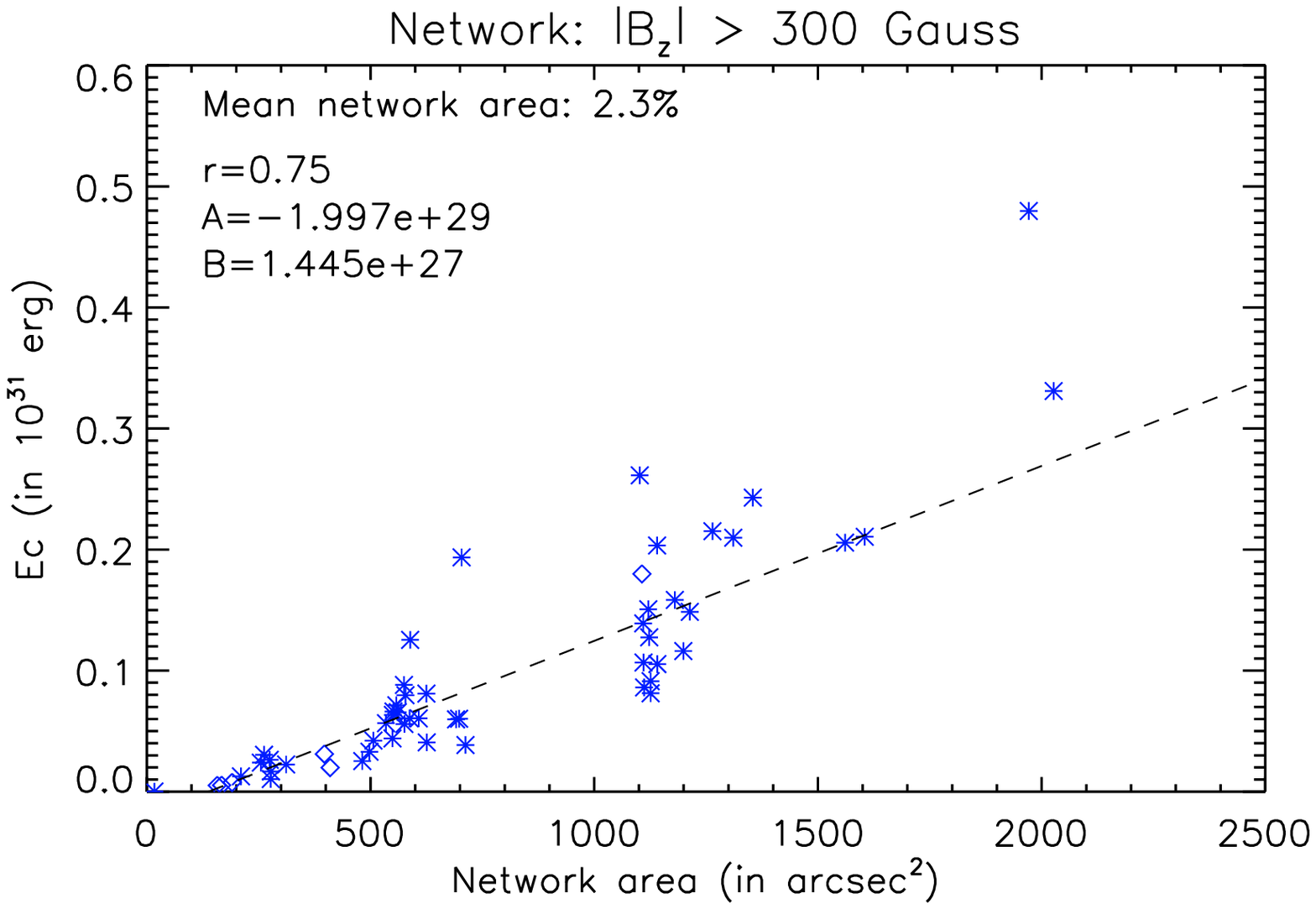}
  \includegraphics[width=0.49\linewidth]{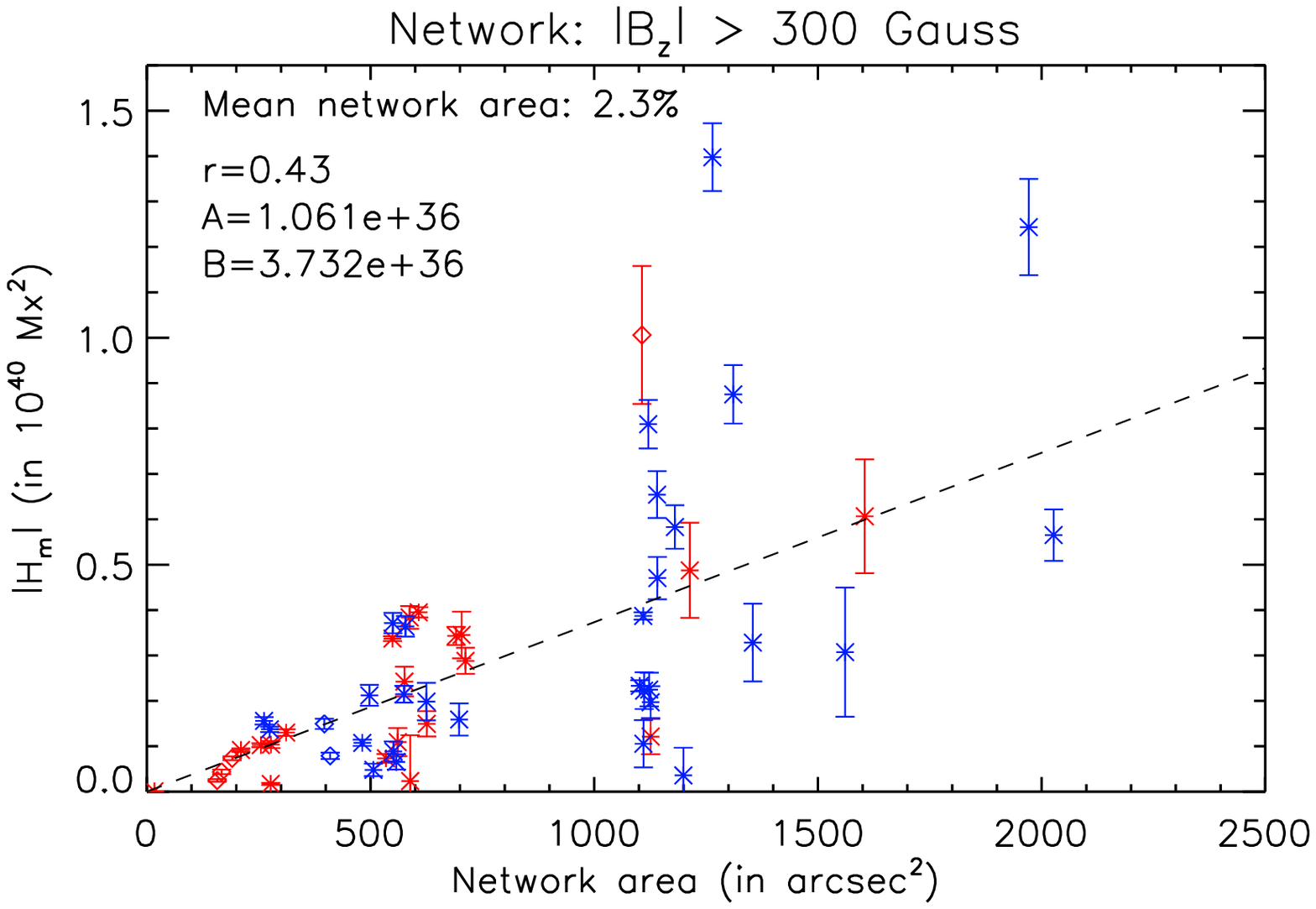}\\
     \caption{Derived free magnetic energy (left column) and relative magnetic helicity (right column) budgets as a function of the image plane magnetic network area, for three different network area thresholds of 100 G, 200 G and 300 G. In plots involving helicity budgets, red/blue colors
     correspond to negative/positive magnetic-helicity budgets. Quiet Sun regions centered within latitudes of $\pm$30\degr\
     (see Fig.~\ref{noaapos}) are represented by asterisks, while regions centered in larger latitudes are presented by diamonds.
     Error bars in helicity are also shown,while respective energy errors are not indicated due to their very small amplitude ($\sim$10$^{28}$ erg).
     Dashed lines show the best linear fit between helicity/energy and network area ($y$=A+B$x$, where $y$ represents
     helicity/energy and $x$ the network area). Derived fitting coefficients $A$ and $B$, the goodness of the fit $r$, as well as the mean network
     area as a percentage of the magnetogram area is also registered within each plot. Derived linear fit coefficients (A, B) between free magnetic energy and heliographic plane network area are equal to (-2.89$\times10^{29}$, 4.33$\times10^{26}$), (-2.53$\times10^{29}$, 9.39$\times10^{26}$), and (-2.21$\times10^{29}$, 1.42$\times10^{27}$) for the three network thresholds, while for relative magnetic helicity and heliographic plane network area are equal to (-5.06$\times10^{37}$, 7.89$\times10^{35}$), (-5.63$\times10^{37}$, 2.31$\times10^{36}$), and (-6.53$\times10^{35}$, 3.54$\times10^{36}$).}
     \label{hmecarea}
\end{figure*}

\section{Results}
\label{res}

\subsection{Energy and helicity budgets of quiet Sun regions}
\label{enhelbudgets}

Using Eqs.~(\ref{Ec_fin}) and (\ref{Hm_fin}) we have derived the free magnetic energy and relative magnetic
helicity budgets for the selected 55 quiet-Sun vector magnetograms.
Free magnetic energy ranges between 7.5 $\times$ 10$^{26}$ erg and 4.8 $\times$ 10$^{30}$ erg while relative magnetic helicity ranges between
7.53 $\times$ 10$^{36}$ Mx$^2$ and 1.4 $\times$ 10$^{40}$ Mx$^2$ (see Fig.~\ref{hmecarea}). Errors for free magnetic energy are in the range of 1.16 $\times$ 10$^{27}$ erg to 5.33 $\times$ 10$^{28}$ erg with a mean error of 1.91 $\times$ 10$^{28}$ erg, while respective errors for relative magnetic helicity range between 2.81 $\times$ 10$^{36}$ Mx$^2$ and 1.52 $\times$ 10$^{39}$ Mx$^2$ with a mean error of 3.7 $\times$ 10$^{38}$ Mx$^2$. The derived free energy and relative helicity budgets are much lower than the respective budgets of ARs \citep[see][]{tzio12}, as Fig.~\ref{ehdiag} also clearly demonstrates; this will be further discussed in Sect.~\ref{enheld}.

Out of the 55 quiet Sun regions, 31 have a positive (right-handed) helicity budget, while 24 have a negative (left-handed) helicity budget.
Unfortunately, we cannot deduce any north/south hemispheric preference for left/right-handed helicity, as for ARs,
since most of the regions are located close to disc center and cover large areas of both north and south solar hemispheres
(see Fig.~\ref{noaapos}). Our results also indicate that there is no dominant sense of net helicity in quiet Sun
regions: the ratio H$_{\rm pos}$/|H$_{\rm neg}$| between the positive (H$_{\rm pos}$) and the negative (H$_{\rm neg}$) helicity
components of the net helicity H$_{\rm m}$ ranges, with the exception of a single region, between 0.32-2.31 with a mean of 1.06.
The region that shows a clear net negative helicity with a H$_{\rm pos}$/|H$_{\rm neg}$| = 6.14 $\times$ 10$^{-5}$ has the lowest derived
free magnetic energy and (absolute) relative helicity budgets of 7.5 $\times$ 10$^{26}$ erg and 7.53 $\times$ 10$^{36}$ Mx$^2$ respectively.

\begin{table*}
\caption{Derived global instantaneous budgets for quiet-Sun free energy and relative helicity during the maximum and minimum of a solar cycle (see Sect.~\ref{netarea}) for different network thresholds.}
\label{table2}
\centering
\begin{tabular}{c c c c c}
\hline
& \multicolumn{4}{c}{Instantaneous budgets}   \\
  & \multicolumn{2}{c}{Helicity (Mx$^2$)} & \multicolumn{2}{c}{Energy (erg)}  \\
Network mask & Maximum & Minimum & Maximum & Minimum   \\
\hline
|B$_{\rm z}$| $\ge$ 100 G & 5.92$\pm$0.52 $\times$ 10$^{41}$  & 1.48$\pm$0.13 $\times$ 10$^{42}$ & 2.54$\pm$0.22  $\times$ 10$^{32}$ &  6.36$\pm$0.56 $\times$ 10$^{32}$  \\
|B$_{\rm z}$| $\ge$ 200 G & 1.42$\pm$0.05 $\times$ 10$^{42}$  & 3.54$\pm$0.13 $\times$ 10$^{42}$ & 5.62$\pm$0.21  $\times$ 10$^{32}$ &
1.41$\pm$0.05 $\times$ 10$^{33}$ \\
|B$_{\rm z}$| $\ge$ 300 G & 2.16$\pm$0.05 $\times$ 10$^{42}$  & 5.4$\pm$0.13 $\times$ 10$^{42}$ & 8.35$\pm$0.19  $\times$ 10$^{32}$ &
2.09$\pm$0.05 $\times$ 10$^{33}$ \\
\hline
\end{tabular}
\end{table*}

\subsection{Energy and helicity budgets as a function of network area}
\label{netarea}

In the quiet Sun, magnetic fields are mainly concentrated in the magnetic network, which is defined by
the boundaries of supergranular cells. Since magnetic field magnitudes within the magnetic network are larger than
field values within the interior of supergranular cells that define the internetwork, the total network area
can be viewed, as the area where the amplitude of the vertical field component B$_{\rm z}$ is larger than a threshold value.
This threshold value is usually of the order 100--300 G, with the most commonly used threshold value being 200 G.
We note, that for our analysis we prefer to consider areas on the image plane, instead of areas on the heliographic plane, as they are a directly observable and measurable quantity. However, with most of the studied quiet Sun regions located close to the solar disc center, derived areas on the image plane do not differ substantially from those derived on the heliographic plane.
Taking into account that the longitudinal magnetic field sensitivity of the SOT/SP is of the order of 5 G, the respective mean error for the network area percentage coverage is 6.2\%, 3.1\% and 2.1\% for the three
considered network thresholds of |B$_{\rm z}$|$>$100 G, |B$_{\rm z}$|$>$200 G, and |B$_{\rm z}$|$>$300 G.

Figure~\ref{hmecarea} shows the derived free magnetic energy and relative magnetic
helicity budgets of the 55 quiet Sun regions as a function of the image plane magnetic network area, derived for the three considered threshold values. The larger the threshold value the smaller the network area coverage as a percentage of the magnetogram area. As Fig.~\ref{hmecarea} suggests, there seems to be a monotonic dependence of free magnetic energy on the network area with a goodness $r$ of the linear fit ranging between 0.71 and 0.76 depending on the threshold-dependent network mask. A similar monotonic relationship, although not as clear, seems to exist also between relative helicity and network area; the goodness of the fit is low and ranges between 0.43 and 0.45.  Derived fitting coefficients for the linear fits between helicity/energy and network area for all three different network thresholds are included on the plots of Fig.~\ref{hmecarea}. The derived mean absolute deviation for helicity is 1.5 $\times$ 10$^{39}$ Mx$^2$, 1.49 $\times$ 10$^{39}$ Mx$^2$ and 1.5 $\times$ 10$^{39}$ Mx$^2$, respectively, for the used network thresholds of 100 G, 200 G and 300 G, while for energy the respective values are 3.14 $\times$ 10$^{29}$ erg, 2.79 $\times$ 10$^{29}$ erg, and 2.76 $\times$ 10$^{29}$ erg. These values are one order of magnitude larger than the respective mean uncertainties of the relative helicity and free magnetic energy.

\begin{figure}
  \includegraphics[width=\linewidth]{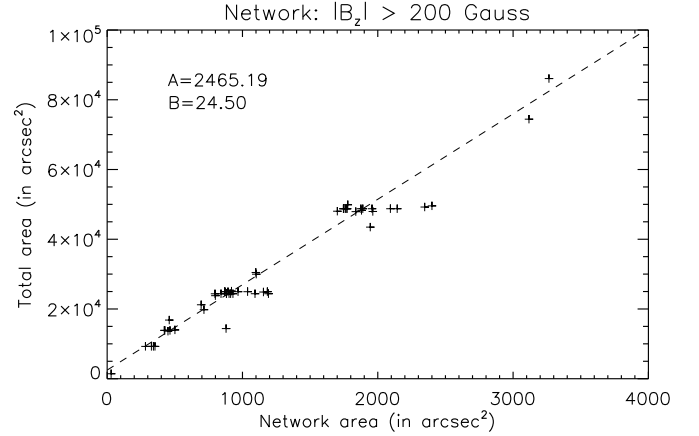}
    \caption{Scatter plot between derived network area and the total area (on the image plane) of the 55 considered quiet Sun magnetograms for a network threshold of |B$_{\rm z}$|$>$200 G. The dashed line shows the best linear fit ($y$=A+B$x$) between the network area ($x$) and the total area ($y$). A similar linear monotonic dependence holds for the other two network thresholds (100G, 300G) considered in this analysis with respective linear fit coefficients (A, B) equal to (2\,177.41, 11.13) and (3\,288.36, 37.9). The linear fit coefficients (A, B) between network and  total area measured on the heliographic plane for the network thresholds (100G, 200G, 300G)  are equal to (2\,337.97, 11.17), (3\,569.72, 23.95), and (3\,463.81, 38.73).}
     \label{areas}
\end{figure}

The aforementioned monotonic dependence of helicity/energy to network area stems from the hierarchical structure of the magnetic field in quiet-Sun
regions that concentrates at the boundaries of supergranular cells. Supergranular cells tend to have rather similar physical characteristics
in terms of sizes and magnetic flux concentrations. As a consequence a) derived helicity and energy budgets roughly depend on the
number of supergranular cells present within the area under investigation and b) the larger the area, the more supergranular cells present and hence
the larger the derived network areas. Fig.~\ref{areas} clearly demonstrates a linear monotonic dependence between network area and the total area of
the quiet-Sun magnetograms.

The network area seems to range from 10\% (maximum phase) to 25\% (minimum phase) of the solar disk area during
a solar cycle \citep[Fig. 3 in][]{cacc98}. Hence, we can derive the instantaneous budgets of free magnetic energy and relative helicity
present {\em globally} in the quiet Sun during the minimum and maximum of the solar cycle using the linear relations shown in Fig.~\ref{hmecarea}, and assuming that the aforementioned network area variance with solar cycle also holds for the rest, non-visible, half solar disc. Since we are dealing with image plane areas, the solar disc is considered to be a circle; all area-dependent results henceforth, if areas on the heliographics plane were  considered instead (whole Sun viewed as a sphere), would be larger by almost a factor of two.
Derived values are presented in Table~\ref{table2}. The instantaneous global budgets of energy for quiet Sun are comparable to the respective budgets of a sizeable AR that is able to produce an eruptive X-class flare. The amplitude of helicity lags behind due to the incoherent helicity sense, resulting in the absence of a dominant helicity sign and hence in even smaller net (algebraic) helicities.

\subsection{Energy--helicity diagram of quiet-Sun regions}
\label{enheld}

\begin{figure}
  \includegraphics[width=\linewidth]{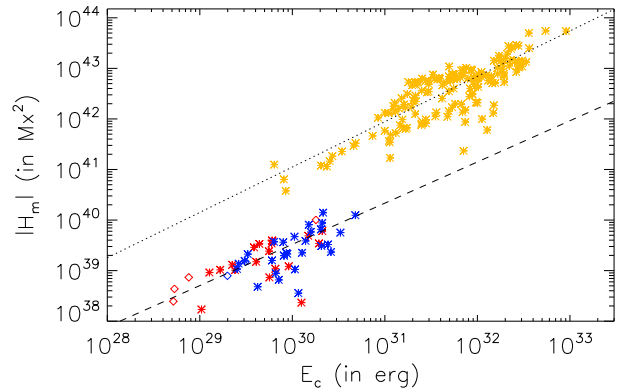}
     \caption{The free magnetic energy - relative helicity diagram of solar quiet regions. Red/blue colors
     correspond to negative/positive total magnetic helicity budgets, while asterisks denote regions centered within heliolatitudes of $\pm$30\degr\
     and diamonds denote regions centered in higher latitudes. The dashed line denotes the least-squares best fit of Eq.~(\ref{best}).
     Yellow asterisks show for comparison the free energy - relative helicity diagram of ARs derived
     by \citet{tzio12}, while the dotted line shows the respective least-squares best fit.
     }
     \label{ehdiag}
\end{figure}

With the derived instantaneous free magnetic energy and relative magnetic
helicity budgets of the selected 55 vector magnetograms we construct, for the first time,
the free-energy--relative helicity diagram (hereafter energy - helicity [EH] diagram) of solar
quiet regions. This diagram is shown in Fig.~\ref{ehdiag}; the respective EH diagram of ARs derived by \cite{tzio12}
is also shown in the same figure for comparison.

Both relative-helicity and free magnetic-energy budgets are much lower in quiet Sun regions than in ARs. This
stems from a) the much lower magnetic fluxes present in quiet Sun regions compared to ARs, b) the less free magnetic energy available in quiet-Sun regions as they are closer to a potential configuration than ARs, and c) the absence of a dominant net helicity in quiet Sun regions, contrary to ARs. As already discussed in Sect.~\ref{enhelbudgets}, the ratio H$_{\rm pos}$/|H$_{\rm neg}$| between the
positive (H$_{\rm pos}$) and the negative (H$_{\rm neg}$) helicity components of the net helicity H$_{\rm m}$, shows a clear incoherence (lack of preference) in helicity sign for quiet-Sun regions.

The EH diagram of Fig.~\ref{ehdiag} shows, like in ARs, a nearly monotonic dependence between E$_{\rm c}$ and H$_{\rm m}$ in quiet-Sun regions.
However, contrary to ARs, this dependence is the result of the presence of hierarchical structures in the network area, as previously
discussed in Sect.~\ref{netarea}, and the monotonic dependence of both free magnetic energy and relative magnetic helicity on the network and total area (Figs.~\ref{hmecarea} and \ref{areas}). This monotonic relation seems to be a logarithmic scaling of the form
\begin{equation}
|H_{\rm m}| \approx 1.153 \times 10^{15} E^{0.815}_{\rm c}
\label{best}
\end{equation}
between absolute relative helicity and free magnetic energy. Let us also note that the EH diagram shows an appreciable segregation between positive-helicity and negative-helicity dominated regions with the former being on the higher end of the EH diagram. Moreover,
the only quiet Sun region showing a net helicity budget (see Sect.~\ref{enhelbudgets}), that possesses both the lowest helicity and energy budgets and is not presented in the EH diagram as they are both almost two orders of magnitude lower than its lower end, also follows the derived linear monotonic relation.

The derived index of 0.815 of the monotonic relation between free magnetic energy and relative helicity in quiet-Sun regions is very close to the respective index of 0.897 for ARs \citep{tzio12} as Fig.~\ref{ehdiag} shows. This suggests that, if quiet-Sun regions had a predominant sign of helicity sign, like in ARs,  their EH diagram would be a nearly continuous extension of the respective diagram for ARs toward lower values. This in fact, is fairly well seen with free magnetic energies only, but helicity extension toward lower values is discontinuous due to the aforementioned incoherence in helicity sign for quiet-Sun regions.

\begin{table}
\caption{Inferred quiet-Sun free energy and relative helicity budgets for an entire solar cycle (see Sect.~\ref{conc} for details) for different network thresholds.}
\label{table3}
\centering
\begin{tabular}{c c c}
\hline
&  \multicolumn{2}{c}{Total within a solar cycle} \\
Network mask  &  Helicity (Mx$^2$) & Energy (erg) \\
\hline
|B$_{\rm z}$| $\ge$ 100 G &
2.1$\pm$1.07 $\times$ 10$^{45}$ & 9.02$\pm$4.58 $\times$ 10$^{35}$ \\
|B$_{\rm z}$| $\ge$ 200 G &
5.02$\pm$2.52 $\times$ 10$^{45}$ & 1.99$\pm$1.0 $\times$ 10$^{36}$\\
|B$_{\rm z}$| $\ge$ 300 G &
7.65$\pm$3.83 $\times$ 10$^{45}$ & 2.96$\pm$1.48 $\times$ 10$^{36}$\\
\hline
\end{tabular}
\end{table}

\section{Discussion and conclusions}
\label{conc}

We have applied the recently proposed NLFF field method by \citet{geo12a} to calculate the instantaneous free magnetic
energy and relative magnetic helicity budgets of quiet Sun regions from single photospheric vector magnetograms.
On a sample of 55 such magnetograms we find, (1) a nearly monotonic relation between free-magnetic-energy/relative-helicity and magnetic
network area (Fig.~\ref{hmecarea}) as well as total area (Fig.~\ref{areas}), and as a consequence (2) a nearly monotonic relation between the free magnetic energy and the relative magnetic helicity in quiet Sun regions (Fig.~\ref{ehdiag}). Derived energy/helicity budgets are much lower than respective budgets of ARs reported by \citet{tzio12}.

Free magnetic energy budgets of quiet-Sun regions represent a rather continuous extension of respective AR budgets towards lower values (Fig.~\ref{ehdiag}). On the other hand, the corresponding helicity transition is discontinuous due to the lack of a dominant sense of relative magnetic helicity in quiet-Sun regions contrary to ARs \citep{tzio12,tzio13}. However, globally quiet-Sun regions show instantaneous budgets of free energy and, to a lesser extent, relative helicity that are comparable to those of a sizable AR. Furthermore, they do not show any hemispheric helicity
preference, contrary to previous reports \citep[e.g.][]{geo09}, but this could well be a selection effect, since most of the analyzed quiet Sun areas cover parts of both north and south hemispheres.

The aforementioned derived monotonic relations between the instantaneous budgets of energy/helicity and magnetic network area
can be used to infer the respective budgets for an entire solar cycle. For such
a derivation the respective helicity/energy injection rates have to be evaluated. Since magnetic flux concentrations used for the calculation of the instantaneous budgets are concentrated in supergranular
boundaries (magnetic network) it can be assumed that the instantaneous budgets of energy/helicity replenish within the lifetime of supergranules, which is of the order of 1.8$\pm$0.9 d \citep[see][and references therein]{rieu10}. However, such an assumption, although valid for the free magnetic energy which is always dissipated through resistive processes, such as reconnection, is not always valid for helicity. The latter, if not bodily removed, is roughly conserved during reconnection and can only be transferred to nearby regions via existing large-scale magnetic connections. Assuming that a) both quantities dissipate/replenish within the aforementioned supergranular lifetime, and b) a sinusoidal variation of the network area between 10\% and 25\% within a solar cycle \citep[see Fig.3 in][]{cacc98} we can derive the total quiet-Sun budgets for free magnetic energy and helicity in a solar cycle. The derived values for different network thresholds are presented in Table~\ref{table3}. The derived helicity budgets of $\sim$10$^{45}$ Mx$^2$ are two orders of magnitude higher than the value of $\sim$10$^{43}$ Mx$^2$ reported by \citet{wels03} and an order of magnitude higher than the value of $\sim$1.5 $\times$ 10$^{44}$ Mx$^2$ reported by \citet{geo09}. However, we must stress the high uncertainties attached to our solar-cycle helicity derivation which mostly stem from the weakness of the used linear fit (Fig.~\ref{hmecarea}) as expressed by the low values of the its goodness $r$. Moreover, the estimates of \citet{wels03} and \citet{geo09} rely on typical photospheric flow velocities and the respective helicity injection rates. These have been found to underestimate the respective integrated budgets \citep{tzio13}.

This work presents the first inference of the quiet-Sun free-energy budget over an entire solar cycle. However, we do know that energy in the magnetic network is dissipated, mostly through reconnection, in fine-scale structures residing at the supergranular boundaries, such as mottles and spicules \citep{tsir:12}.
\cite{tsir04} have reported a value of at least 1.2 $\times$ 10$^5$ erg~cm$^{-2}$~s$^{-1}$ for the energy flux from mottles, while \citet{moor11} reported a value of 7 $\times$ 10$^5$ erg~cm$^{-2}$~s$^{-1}$ from spicules by including co-generated Alv\'{e}n waves.
Assuming again a sinusoidal variation of the network area between 10\% and 25\% within a solar cycle we can integrate the aforementioned values to derive respective energies within a solar cycle of 2 $\times$ 10$^{35}$ erg and 1.2 $\times$ 10$^{36}$ erg. These values are of the same order of magnitude as the derived values of free magnetic energy $\sim$10$^{36}$ erg (see Table~\ref{table3}). Hence, there seems to be enough free energy in the quiet Sun within a solar cycle to power fine-scale structures. \cite{tsir04} have argued that considerable amounts of energy are also needed for heating the chromosphere. We should, however, note that, as discussed in \cite{geo12a}, the derived free magnetic energy is a lower limit and hence there could be even larger amount of free magnetic energy available in quiet Sun regions.

Unfortunately, there exist no estimations of helicity in fine-scale structures, such as mottles and spicules, however such structures often show a helical behaviour \citep{jess09,curd11,depo12,tsir:12}. Whether this behaviour is a manifestation of episodes of helicity removal and how this process actually takes place is still unknown. There exist, however, simulations of larger-scale solar polar jets \citep{pari09}, observed in polar coronal holes, that investigate the reconnection-driven dynamics and the energy and helicity evolution. Future magneto-hydrodynamic (MHD) simulations of fine-scale structures, combined with high resolution observations of the chromosphere, could probably shed light on the processes of heliospheric helicity expulsion - if any - from quiet-Sun regions.

\begin{acknowledgements}
Hinode is a Japanese mission developed and launched by ISAS/JAXA, with NAOJ as domestic partner
and NASA and STFC (UK) as international partners and operated in
co-operation with ESA and NSC (Norway). The Dutch Open Telescope is operated at the Spanish Observatorio del Roque de los Muchachos of the Instituto de Astrof\'{i}sica de Canarias. This research has been carried out in the framework of the Research Projects hosted by the RCAAM of the Academy of Athens. This work was partially supported from the EU's Seventh Framework Programme under grant agreement
n$^o$ PIRG07-GA-2010-268245.
\end{acknowledgements}


\begin{thebibliography}{}

\bibitem[Berger(1984)]{berg84}
Berger, M.~A.\ 1984, Ph.D.~Thesis

\bibitem[Berger \& Field(1984)]{berfie84}
Berger, M.~A., \& Field, G.~B.\ 1984, Journal of Fluid Mechanics, 147, 133

\bibitem[Berger(1999)]{berg99}
Berger, M.~A.\ 1999, Plasma Physics and Controlled Fusion, 41, 167

\bibitem[Caccin et al.(1998)]{cacc98}
Caccin, B., Ermolli, I., Fofi, M., \& Sambuco, A.~M.\ 1998, \solphys, 177, 295

\bibitem[Chesny et al.(2013)]{ches13}
Chesny, D.~L., Oluseyi, H.~M., \& Orange, N.~B.\ 2013, \apjl, 778, L17

\bibitem[Curdt \& Tian(2011)]{curd11} Curdt, W., \& Tian, H.\ 2011, \aap, 532, L9

\bibitem[Demoulin et al.(2006)]{dem06}
Demoulin, P., Pariat, E., \& Berger, M. A. 2006, \solphys, 233, 3

\bibitem[De Pontieu et al.(2012)]{depo12}
De Pontieu, B., Carlsson, M., Rouppe van der Voort, L.~H.~M., et al.\ 2012, \apjl, 752, L12

\bibitem[DeVore(2000)]{devo:00}
DeVore, C.~R. 2000, \apj, 539, 944

\bibitem[Finn \& Antonsen(1985)]{finn85}
Finn, J.~M., \& Antonsen, T.~M., Jr.\ 1985, Commun. Plasma Phys. Controlled Fusion, 9, 111

\bibitem[Galsgaard et al.(2000)]{gals00} Galsgaard, K., Parnell, C.~E., \& Blaizot, J.\ 2000, \aap, 362, 395

\bibitem[Gary \& Hagyard (1990)]{gary90}
Gary, G. A., \& Hagyard, M. J. 1990, \solphys, 126, 21

\bibitem[Georgoulis(2005)]{geo05}
Georgoulis, M. K. 2005, \apj, 629, L69

\bibitem[Georgoulis \& Rust(2007)]{geo:rus}
Georgoulis, M. K., \& Rust, D. M. 2007, \apj, 661, L109

\bibitem[Georgoulis et al.(2009)]{geo09}
Georgoulis, M.~K., Rust, D.~M., Pevtsov, A.~A., Bernasconi, P.~N.,
\& Kuzanyan, K.~M.\ 2009, \apjl, 705, L48

\bibitem[Georgoulis et al.(2012)]{geo12a}
Georgoulis, M. K., Tziotziou, K., \& Raouafi, N.-E. 2012, \apj, 759, 1

\bibitem[Hagenaar et al.(1997)]{hage:97}
Hagenaar, H.~J., Schrijver, C.~J., \& Title, A.~M.\ 1997, \apj, 481, 988

\bibitem[Jess et al.(2009)]{jess09}
Jess, D.~B., Mathioudakis, M., Erd{\'e}lyi, R., et al.\ 2009, Science, 323, 1582

\bibitem[Kontogiannis et al.(2010)]{konto10}
Kontogiannis, I., Tsiropoula, G., Tziotziou, K., \& Georgoulis, M.~K.\ 2010, \aap, 524, A12

\bibitem[Kontogiannis et al.(2011)]{konto11}
Kontogiannis, I., Tsiropoula, G., \& Tziotziou, K.\ 2011, \aap, 531, A66

\bibitem[Leka et al.(1996)]{leka:96}
Leka, K.~D., Canfield, R.~C., McClymont, A.~N., \& van Driel-Gesztelyi, L.\ 1996, \apj, 462, 547

\bibitem[Lites et al.(2008)]{lite08}
Lites, B.~W., Kubo, M., Socas-Navarro, H., et al.\ 2008, \apj, 672, 1237

\bibitem[Liu et al.(2011)]{liu11} Liu, S., Zhang, H.~Q.,
\& Su, J.~T.\ 2011, \solphys, 270, 89

\bibitem[Low(1994)]{low94}
Low, B.~C.\ 1994, Physics of Plasmas, 1, 1684

\bibitem[Metcalf et al.(2006)]{metc06}
Metcalf, T.~R., Leka, K.~D., Barnes, G., et al.\ 2006, \solphys, 237, 267

\bibitem[Metcalf et al.(2008)]{metc08}
Metcalf, T.~R., De Rosa, M.~L., Schrijver, C.~J., et al.\ 2008, \solphys, 247, 269

\bibitem[Meyer et al.(2013)]{meyer13} Meyer, K.~A., Sabol, J.,
Mackay, D.~H., \& van Ballegooijen, A.~A.\ 2013, \apjl, 770, L18

\bibitem[Moore et al.(2011)]{moor11}
Moore, R.~L., Sterling, A.~C., Cirtain, J.~W., \& Falconer, D.~A.\ 2011, \apjl, 731, L18

\bibitem[Pariat et al.(2009)]{pari09}
Pariat, E., Antiochos, S.~K., \& DeVore, C.~R.\ 2009, \apj, 691, 61

\bibitem[Parnell(2001)]{parn:01}
Parnell, C.~E.\ 2001, \solphys, 200, 23

\bibitem[Rieutord \& Rincon(2010)]{rieu10}
Rieutord, M., \& Rincon, F.\ 2010, Living Reviews in Solar Physics, 7, 2

\bibitem[Schrijver \& Harvey(1994)]{schr:harv}
Schrijver, C.~J., \& Harvey, K.~L. 1994, \solphys, 150, 1

\bibitem[Schrijver et al.(1997)]{schr97}
Schrijver, C.~J., Title, A.~M., van Ballegooijen, A.~A., Hagenaar, H.~J.,
\& Shine, R.~A.\ 1997, \apj, 487, 424

\bibitem[Schrijver et al.(2006)]{schr06}
Schrijver, C.~J., De Rosa, M.~L., Metcalf, T.~R., et al.\ 2006, \solphys, 235, 161

\bibitem[Tsiropoula \& Tziotziou (2004)]{tsir04}
Tsiropoula, G., \& Tziotziou, K. 2004, A\&A, 424, 279

\bibitem[Tsiropoula et al.(2012)]{tsir:12}
Tsiropoula, G., Tziotziou, K., Kontogiannis, I., et al.\ 2012, \ssr, 169, 181

\bibitem[Tziotziou et al.(2012)]{tzio12}
Tziotziou, K., Georgoulis, M. K., \& Raouafi, N.-E. 2012, \apj, 759, L4

\bibitem[Tziotziou et al.(2013)]{tzio13}
Tziotziou, K., Georgoulis, M.~K., \& Liu, Y.\ 2013, \apj, 772, 115

\bibitem[Uritsky et al.(2013)]{urit13}
Uritsky, V.~M., Davila, J.~M., Ofman, L., \& Coyner, A.~J.\ 2013, \apj, 769, 62

\bibitem[Wang et al.(1996)]{wang96}
Wang, H., Tang, F., Zirin, H., \& Wang, J.\ 1996, \solphys, 165, 223

\bibitem[Welsch \& Longcope(2003)]{wels03}
Welsch, B.~T., \& Longcope, D.~W.\ 2003, \apj, 588, 620

\bibitem[Welsch et al.(2007)]{wels07}
Welsch, B.~T., Abbett, W.~P., De Rosa, M.~L., et al.\ 2007, \apj, 670, 1434

\bibitem[Wiegelmann et al.(2013)]{wieg13}
Wiegelmann, T., Solanki, S.~K., Borrero, J.~M., et al.\ 2013, \solphys, 283, 253

\bibitem[Woodard \& Chae(1999)]{wood99}
Woodard, M.~F., \& Chae, J.\ 1999, \solphys, 184, 239

\bibitem[Zhao et al.(2009)]{zhao09} Zhao, M., Wang, J.-X.,
Jin, C.-L., \& Zhou, G.-P.\ 2009, Research in Astronomy and Astrophysics, 9, 933

\bibitem[Zirin \& Wang(1993)]{ziri:93}
Zirin, H., \& Wang, H.\ 1993, \nat, 363, 426

\end{thebibliography}
\end{document}